\documentstyle[12pt]{article}

\begin{document}

\title {Global Bifurcations in Rayleigh-B\'{e}nard Convection: Experiments,
 Empirical Maps and Numerical Bifurcation Analysis}
\author {I. G. Kevrekidis and R. Rico-Mart\'{\i}nez\\
Department of Chemical Engineering, Princeton University,\\
Princeton, NJ 08544-5263\\
\and R. E. Ecke$^1$, R. M. Farber$^2$ and A. S. Lapedes$^2$\\
$^1$Physics Division and Center for Nonlinear Studies\\
$^2$Theoretical Division\\
Los Alamos National Laboratory, Los Alamos, NM 87545\\
\ }
\maketitle
\begin{abstract}
We use nonlinear signal processing  techniques, based on artificial
neural networks, to construct an empirical
mapping from experimental Rayleigh-B\'{e}nard
convection data in the quasiperiodic regime.
The data, in the form of a one-parameter sequence of
Poincar\'{e} sections in the interior
of a mode-locked region (resonance horn),
are indicative of a complicated interplay of local and globalbifurcations
with respect to the experimentally varied Rayleigh number.
The dynamic phenomena apparent in the data include period doublings,
complex intermittent behavior,
secondary Hopf bifurcations, and chaotic dynamics.
We use the fitted map to reconstruct the experimental dynamics and
to explore the associated  local and global bifurcation structures
in phase space.
Using numerical bifurcation techniques we locate the stable and unstable
periodic solutions, calculate eigenvalues, approximate
invariant manifolds of saddle type solutions and identify bifurcation
points. This approach constitutes a promising data post-processing
procedure for investigating phase space and parameter space of
real experimental systems;
it allows us to infer phase space structures which the
experiments can only probe with limited
measurement precision and  only at a discrete number of operating
parameter settings.
\end{abstract}
\newpage

\section{Introduction}

When presented with experimental observations of the dynamics of a nonlinear
system at a discrete number of operating parameter settings,
it is often difficult to interpret the global
phase space  structure underlying
the experimental time series, or
to get a good picture of the state of the system at intermediate parameter
values.
It would therefore be desirable to develop a
model of the experimental dynamics that could be implemented, simulated
and analyzed on the computer. Several approaches are possible:
one could start from the
fundamental equations for the system, impose realistic boundary
conditions, and simulate the resulting set of equations. For
experiments described by partial differential equations, such an
approach is often untenable, requiring extensive computational
resources. Even though necessary for the fundamental understanding of
the instabilities underlying the dynamics, this approach
may not {\it quantitatively} reproduce the finer
details of the observed behavior. This may be because of
small deviations from the idealizations of the model,
imperfections of the apparatus, oversimplified boundary conditions,
imperfect estimation of physical properties, inaccurate readings
of the experimental settings, etc.
The second approach, which more practically aims at interpreting the
{\it particular experimental observations}, is to fit the observed dynamics
with a dynamical system (a map, or a set of ODEs) of the appropriate phase
space dimension.
As we shall see, this approach is capable, at least for the
experimental data we have considered, of reproducing significant
features of the dynamics and yielding  insight into the
properties of the phase-space structure underlying the observed behavior.
Techniques for the reconstruction of phase-space mappings have
been developed for dynamical systems analysis \cite{fraser,packard} and
recently for the purpose
of prediction and forecasting of time series(see for example \cite{Fore_rev}
and references therein).
These methods
include functional approximation using radial basis functions,
local-linear maps, and neural network algorithms.

In this paper we use the latter approach (i.e., fitting the observed
dynamics) incorporating in the model a functional dependence on
a bifurcation parameter; this can easily be generalized to
multiple parameters. This type of empirical modeling
provides crucial assistance in conjunction with numerical bifurcation
analysis \cite{bryanjef,kevrekidisNBT} for interpreting
the bifurcation structures ``contained in"
(or consistent with) the experimental data.
More generally, such techniques can play a significant role in real-time
prediction and control applications for a large class of physical systems
(see for example the recent experimental literature on stabilizing unstable
periodic orbits \cite{schwar,yorkeco}).
They can be thought of as a useful data compression and post-processing
tool, which assists the concise presentation and
interpretation of experimental time series.

In this work we consider the
dynamics of data from a Rayleigh-B\'enard convection
experiment in a region of parameter space where the behavior is complicated
and difficult to interpret by visual inspection of the Poincar\'e section
data alone. The paper is organized as follows:  the
next section is a description of the experiment and of the experimental
data. We then outline briefly the data processing
technique used (Artificial Neural Networks, ANNs)
and discuss its performance in fitting the data. Using the fitted map(s) we
construct a plausible picture of the bifurcation structures in phase and
parameter space using numerical bifurcation methods.
Finally, we discuss the applicability of this approach to a wider class
of problems.

\section{Experimental Data}
The experimental data presented here are obtained from a
rectangular small-aspect-ratio
convection cell used for studies of low-dimensional nonlinear dynamics.
The convecting fluid is a 1.46 \% solution of $^3$He in
superfluid $^4$He and thermal convection is induced
by applying a fixed heating current to the top plate while
maintaining the bottom plate at fixed temperature.
Details of the experimental apparatus and properties of the superfluid
mixture are described elsewhere \cite{eckePRA,maenoJLTP}. Here we
discuss the measurement probe and present a brief characterization of the
parameter space of thermal convection in this system.

For a fixed geometry, two dimensionless parameters
characterize the convective state.
The Rayleigh number R, proportional to the
temperature difference across the fluid layer $\Delta T$, is a
measure of the driving force applied to the system.
The other dimensionless parameter in the problem is the Prandtl
number $\sigma$, which measures the ratio of the rates of molecular
diffusion of momentum and heat and controls details of the secondary
instabilities in thermal convection.  In $^3$He-$^4$He mixtures
$\sigma$ is a strong function of the mean temperature T$_m$ of the
cell, and is varied over the quasiperiodic regime in the range
0.06$ < \sigma <$ 0.08
by varying the mean temperature in the range $0.83 < T_m < 0.90
 {\rm K}$.

For small R, the fluid conducts heat diffusively. At
$R \approx 2000$ there is a forward (supercritical) bifurcation to steady
convection \cite{maenoJLTP} where the fluid motion is believed to be
two convection rolls oriented parallel to the short side of the
rectangular convection cell.
At higher values of R the flow becomes
time dependent, going through a sequence of bifurcations from
periodic to quasiperiodic to chaotic dynamics.
The state of the time-dependent convective flow is partially
determined from measurements of the spatially-averaged
convective heat transport.
In small-aspect-ratio convection, however,
the side walls severely constrain the
spatial structure of the fluid flow, and measurement of the dynamics of
the system at a single point is often adequate to characterize the
dynamical nature of the state. Therefore,
in addition to a global measure of the temperature field from
the time dependence of the top-bottom temperature difference, we
use a thermal probe which measures
a local temperature difference near the
center of the cell top plate with a temperature sensitivity of
$0.3 \times 10^{-7} {\rm K}/ {\rm Hz^{1/2}}$ \cite{maenoJLTP}.
We denote the measured
temperature difference at the local probe by $\delta T(t)$.
The output of this probe is digitized to produce a time series record
of the system dynamics.

\begin{figure}[t]
%\centerline{\psfig{file=./figures/pdiag2.ps,height=2.5in,width=3.2in}}
\vspace{2.5in}
\caption{Convective dynamic states in two-parameter
(Rayleigh number and Prandtl number) space.
Dynamic behavior in the various regions is described in the text.}
\label{fig:pdiag}
\end{figure}

In order to understand the context of the data, we now describe
briefly the relevant regions of parameter space, see Fig. \ref{fig:pdiag}.
The first time dependent solution ``begins" at the transition to periodic
oscillations of frequency $f_1$ (typically $f_1 \approx$ 0.6 Hz).
This transition is a forward Hopf
bifurcation in R  \cite {eckePRA}, and its critical R value
depends strongly on $\sigma$. Another Hopf bifurcation  at higher R
gives rise to a second frequency $f_2$, incommensurate with the first.
This second mode is only weakly interacting with the initial limit
cycle mode and not until there is a discontinuous transition to a
different second mode does measurable mode-locking occur.
Within a region of parameter space above the discontinuous
transition, quasiperiodic (incommensurate frequencies), mode-locked
and chaotic states exist \cite{eckePLA,eckePRA2,hauckePHYSD}.

\begin{figure}[p]
%\centerline{\psfig{file=./figures/eckeplot6ad2.ps,height=5.3in,width=5.5in}}
\vspace{5.3in}
\caption{The various data processing stages of the experimental run:
(a) a segment of the experimental time series, (b) the power spectrum,
(c) a projection of the reconstructed attractor using time delays (notice
the location of the plane (x(t)= -0.15) used to
generate the Poincar\'e section  shown in (d)).}
\label{fig:comp4ecke}
\end{figure}

Fig. \ref{fig:comp4ecke}
shows the various data processing stages of a ``typical"
experimental run, in this case for $R/R_c$ =12.01, 1/$\sigma$ = 14.97.
(a) shows a segment of the experimental time series;
(b) shows the corresponding
power spectrum (calculated from the time series using an FFT
algorithm with standard windowing techniques),
marking the frequencies $f_1$ and $f_2$ from fitting.
Standard techniques of phase space reconstruction \cite{fraser,packard} are
used to produce phase space trajectories of the dynamical system;
(c) shows a projection of the continuous attractor reconstructed with
two time delays, $\tau_1 = 2 s$ and $\tau_2 = 1 s$ (the phase space
for all the attractors in this paper has been normalized so that the
limits of the attractors are $\pm$1); and finally (d)
shows the Poincar\'e section obtained from this attractor by transversely
intersecting it with a plane defined by x(t)= -0.15 and only plotting
intersections in a single direction. How this particular section arises
will be discussed later. The location
and orientation of the Poincar\'e plane is variable, so that we can choose
sections with the minimum of overlappings and projection
singularities usually arising in delay coordinate
reconstructions of dynamical attractors.
Using these techniques we can characterize
the evolution of dynamical states of the system as R is varied.

The dynamics in the quasiperiodic regime are
often ``summarized" in terms of a single number, the winding or
rotation number
$W\equiv f_2/f_1$, where $f_1$ and $f_2$ are the fundamental
frequencies (practically determined from spectral analysis of the
time series data).
The mode-locking
structure known as the ``devil's staircase" is constructed from a sequence
of such spectra as a function of $R/R_c$ at fixed $\sigma$.
Such a representative structure is illustrated in Fig.~\ref{fig:exstair},
obtained from experimental data (the Rayleigh number is normalized
with the critical $R_c=2000$). By making a series of
such measurements at different values of $\sigma$, the locking regions
in the \{$R/R_c$,$\sigma$\} parameter space are determined,
Fig. \ref{fig:extongues}. These regions, called ``Arnol'd horns'' or
``resonance" horns,
are seen to broaden as $\sigma$ decreases (this is why we plot
$1/\sigma$ in the ordinate in Fig.~\ref{fig:extongues})
indicating that the coupling or
nonlinearity of the oscillatory modes varies roughly with $1/\sigma$.
Similarly the value of the winding number is
controlled primarily by changing $R/R_c$.

\begin{figure}[p]
%\centerline{\psfig{file=./figures/stair.ps,height=2.5in,width=3.6in}}
\vspace{2.5in}
\caption{Frequency ratio f$_2$/f$_1$ vs. R/R$_c$ showing
experimental devil's staircase of mode-locked intervals for
1/{$\sigma$}=14.9. Prominent lockings are
indicated.}
\label{fig:exstair}
\end{figure}

\begin{figure}[p]
%\centerline{\psfig{figure=./figures/tong.ps,height=15pc,width=30pc}}
\vspace{15pc}
\caption{Experimental regions of mode locking in the 1/$\sigma$
and R/R$_c$
parameter space. The dashed lines represent the hysteretic
discontinuity at low R/R$_c$, increasing R/R$_c$ (- - - - -) and
decreasing (-- -- -- --), and a transition to high dimensional
chaos at large R/R$_c$. For 1/$\sigma\stackrel{>}{~}$15
there is structure in the horns not shown on this global plot.
}
\label{fig:extongues}
\end{figure}

\begin{figure}[t]
\vspace{3in}
% No postscript available for this figure
\caption{Fine structure of the $2/13$ resonance horn. Secondary Hopf
bifurcations (S.H.B.), period doublings, hysteresis, intermittency and
resonance horn overlaps are shown.}
\label{fig:extongues2}
\end{figure}

\begin{figure}[t]
%\centerline{\psfig{file=./figures/wind2splot.ps,height=3.15in,width=3.2in}}
\vspace{3.15in}
\caption{W vs R/R$_c$ for ascending (+) and descending ($\circ$)
sequences ($1/\sigma=14.97$).
Prominent lockings are labeled.}
\label{fig:wind2}
\end{figure}

For $1/\sigma < 14$, the dynamics of the system are well described by
the circle map model \cite{ecke_nato,eckePRA2,hauckePHYSD}. For
special irrational winding numbers, the universality predicted for the
attractor has been verified in this
system \cite{eckePRA2,mainieriPRL1}. Somewhat higher up in the
horns more complicated behavior appears, arising via global
bifurcations \cite{eckePLA,kevrekidisCM}. It is this region
of parameter space that concerns us here. In Fig.~\ref{fig:extongues2}
we illustrate the detailed structure in the
neighborhood of the W=2/13 locking.
Significant features in addition to quasiperiodic and mode-locked
regions include secondary Hopf bifurcations of the mode-locked
periodic orbit, period doubling bifurcations, regions of complicated
transient behavior,
and apparently chaotic states whose appearance is probably associated with
the occurrence of global bifurcations.
Hysteresis in the transitions between these states can be thought of as
indicative of such global bifurcations, as computational and
theoretical work on many model systems shows. Fig.~\ref{fig:wind2} is
a representative experimental plot of the hysteretic behavior
of the winding number for ascending and descending sequences of $R/R_c$
\cite{kevrekidisCM}.
The effective winding number of the Poincar\'e sections, $\rho$, is
related to W measured from the power spectra by $\rho$ = 1/W, mod 1. So
for W=2/13 we get $\rho$ = 1/2 which gives a period-2 cycle with two
points in the Poincar\'e section \cite{ecke_nato,hauckePHYSD}. Our
purpose in the next sections is to describe the complicated dynamics
represented by these data with a simple empirical mapping as a
function of a single control parameter, $R/R_c$. We will use fitting
techniques to generate such an input-output map that can be analyzed with
numerical bifurcation methods.

\section{Empirical Map Construction}
We used a standard ANN configuration \cite{lapedes_farber} to
process the time series from the experimental results sampled
in Fig.~\ref{fig:ps2_sequence}. The experimental information
is in the form of a (nonlinear) map $F$:
$$ {\bf X_{n}} \stackrel{{\bf F}({\bf X},R)}{\longrightarrow}{\bf X_{n+1}}$$
\noindent
where $ {\bf X} = [x \quad y]^T $ are
the coordinates of the intersections of the
trajectory with the Poincar\'e plane, $ {\bf X_n}$ and  $ {\bf X_{n+1}}$ are
two such successive intersections, and R is the value of the
operating parameter, the Rayleigh number.

\begin{figure}[p]
%\centerline{\psfig{file=./figures/figure7.eps,height=5in,width=2.5in}}
\vspace{5in}
\caption{A sampling of the experimental Poincar\'e sections used for training
the ANN. The experimental observations depicted in the ``upper" sequence
(O through L) are indicative of a large amplitude invariant circle
while the ones in the ``lower" sequence (A through J)
are related to the period-2 resonant solution.
The contiguous sequence A through J was broken
in the figure only because of space limitations.
The overlap zone demonstrates the observed bistability (hysteresis).
The parameter values ($R/R_c$) are as
follows A: 12.057, B: 12.040, C: 12.032, D: 12.019, E: 12.0150,
F: 12.011, G: 12.007, H: 12.002, I: 11.986, J: 11.982, K: 11.987,
L: 11.983, M: 11.979, N: 11.950 and O: 11.946.}
\label{fig:ps2_sequence}
\end{figure}

While a first visual inspection of the phase portraits in
Fig.~\ref{fig:ps2_sequence} might indicate that a map of the plane would
constitute a satisfactory representation of the data, a more careful
study reveals that there exist folds in the projection used
(see for example
Fig.~\ref{fig:ps2_sequence}H or Fig.~\ref{fig:ps2_sequence}L).
We were not able to find a two-dimensional plane that would ``get rid
of" these projection singularities for {\it all} the phase portraits
involved in this one-parameter cut. For that reason, and in order
to obtain a deterministic map, we used one more delay
in the reconstruction of the attractor. This means that we chose to
fit a (four-dimensional) map ${\bf F}$ of the form
$ {\bf X_{n+1}} = {\bf F} ({\bf X_n,
X_{n-1}}, R) $.

\begin{figure}[t]
%\centerline{\psfig{file=./figures/eckenet2.ps,height=3.5in,width=6in}}
\vspace{3.5in}
\caption{ Schematic representation of the ANN configuration used.
All layers are
fully interconnected; a few of the connections are depicted.}
\label{fig:netschem}
\end{figure}

Artificial neural networks (see Fig. 8)
are structures composed of many interconnected
processing units (neurons). A fully connected network distributes the
outputs of every neuron in a given
layer to all the neurons of the layer above.
The input layer is composed of ``fan-out" units whose function is only
to distribute their inputs to the neurons in the
next layer.

The neurons of the intermediate or ``hidden" layers compute as
their output
a scalar {\it nonlinear} function (usually of sigmoidal shape) of a
weighted sum of
their inputs. The input to one of the neurons of the first hidden layer
is the sum \( \displaystyle{\sum_{i}}a_iX_i+b \); $X_i$ are
the outputs of the layer below (the input layer), and the constants
$a_i$ and $b$ (which are different for each neuron and are called
``weights" and ``offsets", respectively) will be determined, as we
discuss below, by ``training" the network. The outputs of the neurons
of this first hidden layer will then serve as inputs of the neurons of
the second hidden layer, with new weights and offsets, and so on.
The neurons of the output (final) layer simply produce {\it linear}
functions (linear combinations plus a scalar offset) of their inputs.

These structures have been found to have universal
approximator properties: they can be used to construct
approximations (outputs) of continuous functions of n real
variables (inputs) with support in the unit
hypercube \cite{cyb,hor}. The standard ANN
architecture used in this work consists
of a four layered network: input layer, output
layer and two hidden layers.
This particular structure has been found to be, in
practice, successful in the identification of nonlinear
mappings based on time series data (e.g. \cite{lapedes_farber}).

The feedforward artificial neural networks we used had
an input layer consisting of four
neurons for the system state and one neuron for the operating parameter.
The output layer consisted of two neurons that predict
the point at which the continuous
time trajectory will next intersect the plane which defines the
Poincar\'e section. In order to successively
iterate the ANN, both outputs have
to be fed back into the corresponding inputs
(schematically depicted in Fig.~\ref{fig:netschem} with dash--dot lines).

The weights and offsets of the ANN
that provide the ``best" approximation
(outputs) of the observed measurements as a function of
previous measurements (inputs) are determined by ``training"
the network:
a least squares minimization problem. The objective function
is the norm of the
difference between ANN predictions and actual experimental
measurements of
the states after one iteration of the Poincar\'e map.
Training was performed using a conjugate gradient algorithm, with the
map and derivative evaluations performed in parallel using a
SIMD computer (the 64,000 processor CM-200) at the
Advanced Computing
Laboratory at LANL. The implementation we used is therefore
ideally suited to
very large sets (many thousands) of data points.

In the process of this research we trained several ANNs, all of them
containing two hidden layers; the ``best" results presented
here were obtained
with 10 nonlinear neurons per hidden layer, each neuron with activation
function $g(X)={{1}\over{2}}(1+tanh(X))$,
in addition to the input  and  output layers whose neurons are linear
(see Fig.~\ref{fig:netschem}).
As there is currently no rigorous way to determine an optimum number
of neurons for a particular set of data, the number of neurons
in the hidden layers is somewhat arbitrary
(see for example \cite{kramer}).
Our choice here reflects a compromise between the
computational effort required
to train the ANN and an estimate of the minimum number of neurons
needed to capture the underlying dynamics.

The training set from each ``experimental run" contains about
300 Poincar\'e section
points; an ``experimental run" consists of time series measurements obtained
at a single value of the operating parameter, and is considered to be
converged on the attractor, in the sense that measurements during an
initial period following the parameter change (transient data approaching the
attractor) are not used in the training.  In principle, it would be
desirable to include such transient data in the training, since this
would provide information  about the dynamics ``away" from
the attractor in phase space,
and would also provide to the ANN quantitative
information about the rate of approach to the attractor
(and thus its stability characteristics).

A total of 55 different experimental runs were available for this value of
the Prandtl number, 1/$\sigma$=15.04
(27 from experiments obtained with {\it increasing} R,
and the remaining  obtained with {\it decreasing} R). Appendix A
(Figures~\ref{fig:all1} and \ref{fig:all2})
at the end of the paper shows the entire set of experimental
observations used for training. For this comparatively small data set
a CM-200 is not necessary to successfully train a network
in a realistic amount of time. However, as the
complexity and the dimensionality
of the behavior grow, along with the size of the data set,
our parallel implementation should become indispensable.

\section{Interpretation of the data}

In what follows, we will present our {\it a priori} ``best guess" of a
consistent sequence of
bifurcations underlying the phase portraits in Fig.~\ref{fig:ps2_sequence}
as illustrated in the schematic
bifurcation diagram of Fig.~\ref{fig:schemebif1}.
While some of our interpretations may appear somewhat arbitrary at first,
a reader experienced in the study of resonance phenomena for maps
or periodically-forced oscillators will find such
bifurcation sequences in the quasiperiodic regime quite familiar.

\begin{figure}[p]
%\centerline{\psfig{file=./figures/rbschem1-2.eps,height=4.5in,width=6in}}
\vspace{4.5in}
\caption{ Tentative schematic bifurcation diagram of the transitions observed
in the experimental data. Solid lines indicate stable periodic solutions,
broken lines indicate unstable solutions and filled circles indicate
period-2 small amplitude invariant circles (small) or large amplitude
invariant circles (large). }
\label{fig:schemebif1}
\end{figure}

\begin{figure} [p]
% This figure not available as postscript.
\vspace{5in}
\caption{ A sequence of Poincar\'e sections obtained at $1/\sigma=14.749$,
lower down the $2/13$ resonance horn. They clearly indicate the birth of
a stable (resonant) period-2 in a saddle-node bifurcation {\it on} the
invariant circle.}
\label{fig:lowcm}
\end{figure}

Our data start at $R/R_c=12.052$, roughly in the middle of the $2/13$
resonance horn; we clearly
see a period-2 attractor, phase portrait A in Fig.~\ref{fig:ps2_sequence}.
We will call this the ``resonant" period-2; it is
associated with the 2/13 resonance horn, and the boundaries of the horn
correspond to saddle-node bifurcations involving this period-2 solution. At
higher values of the Prandtl number (lower down the horn in
Figs.~\ref{fig:extongues} and ~\ref{fig:extongues2}) the data clearly
indicate that this saddle-node bifurcation
occurs on a smooth invariant circle,
Fig.~\ref{fig:lowcm} \cite{kevrekidisCM}.
One therefore expects a saddle period-2 to coexist with
the stable period-2, and a ``minimal" requirement
of the fitted map would be to
predict the existence of such a solution. The fitted map should
also exhibit a
saddle-node bifurcation involving this saddle
and the resonant period-2
at low values of R, i.e. at the boundary of the resonance horn.

A supercritical (soft) Hopf bifurcation occurs with decreasing R,
giving rise to
a small amplitude period-2 invariant circle (phase portrait B).
The model should
be capable of predicting this Hopf bifurcation, i.e.,
the stable period-2 solution
should lose stability with two eigenvalues of its linearization
exiting the unit
circle in the complex plane. Furthermore, the model should predict
an unstable (source)
period-2 solution surrounded by the period-2 stable invariant circle.

The period-2 invariant circles grow in amplitude and deform (phase
portraits C and
D), and the first ``nontrivial" transition occurs between phase portraits D
(showing a period-2  ``cuspy" invariant circle) and phase portrait E, showing
a period-16 stable solution. The period-16 points are obviously located close
to the eight ``cusps" or corners developed by the period-2 circles. It is in
principle possible that these apparent cusps may be the result of a
singularity due to the projection to a two-dimensional picture. Because of
the intensely deformed nature of the circles, however, we do not believe
that this period
16 is the result of frequency-locking  {\it on} the circles.
We suggest that a saddle-node of period-8
solutions occurs {\it away from} the period-2 circles. The period-2 circles,
growing in
amplitude, are then lost via a global bifurcation involving their interaction
with the saddle period-8 solution (this is consistent
with the pronounced cusps
developing on the invariant circles). The stable period-8 solution
then undergoes
a cascade of period doublings, resulting first in the observed period-16
(phase portrait E)
and eventually in an 8-horseshoe-piece apparently chaotic attractor
(phase portrait F).
This attractor then undergoes a reverse sequence of
period doublings, coming back to a period 16 (phase portrait G),
a period 8, and finally resulting in phase portrait H.

Fig.~\ref{fig:schemebif1} summarizes our interpretation of these transitions:
the stable period-2 undergoes a ``soft" Hopf bifurcation, and the resulting
stable period-2 invariant circle grows and is
eventually destroyed in a global bifurcation involving
the stable and unstable manifolds of the saddle-type period-8 solutions. These
period-8 solutions exist on an isola: a saddle-node pair of period-8 solutions
is born and eventually disappears in saddle-node period-8 bifurcations. The
stable node period-8 becomes a focus, then an inverse node, and finally
undergoes
a period-doubling cascade, followed by a reverse cascade, and disappearing in
the ``other" saddle-node end of the isola.  Notice that our interpretation
predicts a small hysteresis interval where stable period-8 solutions coexist
with a stable period-2 invariant circle. It is in principle possible
to avoid this by assuming that the period-8 solution
results from frequency locking {\it on}
the period-2 invariant circles; it is the
``pointed" nature of the shape of the circles that argues against such
an interpretation.

It is tempting to consider that the reverse sequence of these bifurcations
occurs as the bifurcation parameter R is further reduced: A global bifurcation
involving the saddle period-8 stable and unstable
manifolds ``gives rise" again
to large amplitude period-2 invariant circles (slightly wrinkled)
in phase portrait H. The amplitude of the circles
diminishes (phase portraits I and J), but here
the analogy with the high R behavior stops:
there is no evidence of a low-R Hopf bifurcation,
nor is there a stable period-2 observed in the data. Instead,
the behavior ``jumps" directly
from large amplitude period-2 invariant circles to a large amplitude attractor.
Indeed, this large amplitude attractor coexists with the large amplitude
period-2
invariant circle (that is the hysteresis
observed at the edge of the resonance horn
in Fig.~\ref{fig:wind2}). The large amplitude attractor gradually
develops into a smooth large amplitude invariant circle in
phase portrait N, and
shows clearly a saddle-node period-5 frequency locking between
phase portraits N
and O. That this latter bifurcation is indeed a saddle-node bifurcation
{\it on} the invariant circle is supported by the
accumulation of experimental points on the large invariant
circle in phase portrait N,
occurring at five distinct locations. The stable period-5 should be
accompanied by a
saddle period-5, and has rotation number $f_2 / f_1 =  5/33$.
This would then imply
that frequency lockings predicted by the Farey sequence should
exist (and the largest
ones could be detected) between the 5/33 and the 2/13 resonances: there is
experimental evidence of the 12/79 and 7/46 lockings in the data.

The large amplitude invariant circle (denoted by
large filled circles in Fig.~\ref{fig:schemebif1})
develops a number of frequency lockings (indicated by the
``flat" parts of its evolution in the bifurcation diagram),
reminiscent of the truly
flat intervals of frequency locking in a rotation number vs. R plot.

Based on this information, here is a plausible explanation of what occurs
as R grows from lower to higher values: At some point an
``invisible" saddle-node (saddle-source) bifurcation of period-2 solutions
occurs, away in phase space from the large amplitude invariant circle.
A global bifurcation,
involving the saddle-type period-2 solutions gives rise to finite-amplitude
period-2 invariant circles, thus leading to an interval of hysteresis between
the stable period-2 invariant circles and the large amplitude attractor; this
is the hysteresis observed experimentally close to the boundary of the horn.
In this case of multistability, the basin boundary between the two attractors
(the large-amplitude invariant circle and the period-2 invariant circles)
is provided by the stable manifolds of the saddle-type
period-2 solutions. Finally,
these same saddle period-2s are responsible for
the end of the hysteresis interval:
the large amplitude invariant circle approaches these
two saddle points,
and is lost
in a global bifurcation involving their stable and unstable manifolds.

This interpretation is ``realistic" in that it involves
only generic codimension-1
bifurcations, which are furthermore known to occur in similar
regimes for maps of
the plane (see for example \cite{aronson,gambaudo,peckham}.)
The only ``unusual" element is
that the saddle-node bifurcation at the
boundary of the resonance horn is now a saddle-source;
both period-2 solutions
born there are unstable. This interpretation is due to the
fact that no stable
period-2 points were observed for low R in the one-parameter diagram.
We saw that ``lower down" the resonance horn (lower 1/$\sigma$) this
saddle-node bifurcation yields a stable node (Fig.~\ref{fig:lowcm});
this would then imply the existence (in a two-parameter continuation) of
a Takens-Bogdanov point \cite{bogdanov,takens} (two
eigenvalues of the resonant period-2 solution at 1) on
the boundary of the resonance horn somewhere in the vicinity
of our experimental one-parameter cut. Since secondary Hopf bifurcation
curves emanate
from such codimension-2 bifurcation points
\cite{peckham,gambaudo,kevrekidis2}, this interpretation is also
corroborated by the existence of period-2 invariant circles nearby,
and the observation of
secondary Hopf bifurcations along the experimental one-parameter cut.
A plausible embedding of this one-parameter cut in a two-parameter diagram
is presented and discussed in Section 6.

This inferred sequence of bifurcations is consistent with the observed
phase portraits. The important question is whether a model based on -- and
consistent with -- the experimental data will indeed exhibit
these bifurcations
at intermediate values of the control parameter (R).

\begin{figure}[p]
%\centerline{\psfig{file=./figures/ecke1-2bifcomp.ps,height=5.9in,width=6in}}
\vspace{5.9in}
\caption{ Comparison of experimental (top row) versus predicted
(bottom row) attractors, and partial bifurcation diagram predicted by
the first ANN trained. The parameter values ($R/R_c$) for the
portraits are as follows: A: 12.0196, B: 12.033, C: 12.041, D: 12.050.
A' and D' same as A and D, B': 12.0200, C': 12.023.}
\label{fig:eckefig}
\end{figure}

\section{The ANN based models and their predictions}

As we discussed in section 3,
we chose to train an ANN with five neurons in the input
layer, two in the output
layer and two-hidden layers with 10 neurons each. The training set consisted of
data obtained at ascending (Fig.~\ref{fig:all1})
and descending (Fig.~\ref{fig:all2}) values of R; the entire set
of available data was used.
The training was successful, in that the resulting map was capable of
accurately
predicting the one-step-ahead data it was fitted to.
More importantly, when
this map was iterated indefinitely, the qualitative nature and quantitative
location of its attractors was comparable to the experimental ones.
We will start investigating and comparing the numerically predicted and
experimentally observed
(actually inferred) bifurcations
from the interior of the horn $R/R_c \approx 12.049$ towards its left boundary
with decreasing values of $R/R_c$.
Figure ~\ref{fig:eckefig} shows a sequence of qualitative transitions
between experimental phase portraits (top row) and predicted long-term
attractors (bottom row). A bifurcation diagram constructed using the model
is also shown in Fig.~\ref{fig:eckefig}: the stable period-2 indeed undergoes
a Hopf bifurcation at $R/R_c = 12.0257$ (the experimental value is somewhat
higher, at $R/R_c \approx 12.046$). Stable and unstable solutions have been
calculated through Newton iteration and standard continuation techniques.
Local bifurcations were found using standard numerical bifurcation algorithms
(e.g. the package AUTO by Prof. E. Doedel \cite{doedel}).
The model does predict the supercritical
``soft" nature of the bifurcation. The amplitude of the period-2 invariant
circles
grows, and they develop sharp ``corners" as do
the experimental attractors.

A saddle-node of period-8 solutions is observed at $R/R_c = 12.0210$,
and the model
predicts a short interval of hysteresis (coexistence of the stable period-8 and
stable period-2 invariant circles). The period-2 invariant circles are lost
after a global bifurcation involving the stable and unstable manifolds of the
saddle-period 8 solutions. This bifurcation is illustrated in
Fig.~\ref{fig:global}.
The figure shows (enlarged in the vicinity of one of the two period-2 invariant
circles) the situation {\it before} (panel A) and {\it after}
the bifurcation (panel B).
In panel A both sides of the one-dimensional unstable manifold of the
saddles period-8 are eventually attracted by period-8 nodes (no hysteresis).
In panel B one side of the unstable manifold approaches the period-8 nodes,
while the other side spirals inward and asymptotically approaches the
stable period-2 invariant circle.
The change in the structure of the unstable manifolds of the
saddle period-8 solutions
indicates the nature of the bifurcation: a homoclinic crossing
between the stable and
unstable manifolds of the saddle period-8. In this case the
unstable manifold is one-dimensional, while the stable manifold
is three-dimensional
(since we are using a map in ${\cal{R}}^4$). This global bifurcation
does not occur at a single parameter value, but over an interval in
parameter space (starting with a homoclinic tangency, followed by crossing
and ending again by a tangency between the stable and unstable manifolds
of the saddle). This interval is extremely narrow here (almost impossible
to computationally resolve in double precision) and for all practical
purposes this transition occurs at $R/R_c \approx 12.01948$.

\begin{figure}[t]
%\centerline{\psfig{file=./figures/ec1-2pe8global.ps,height=2.625in,width=6in}}
\vspace{2.625in}
\caption{Phase portraits before and after the global bifurcation
associated with the disappearance of the period-2 invariant
circles as predicted by the
first ANN trained. The parameter values ($R/R_c$) are
A: 12.0193 and B: 12.0196. $\bullet$ indicates saddles and $o$ indicates
nodes. Notice the difference in the relative location of the saddle
unstable manifolds in the two pictures.}
\label{fig:global}
\end{figure}

The qualitative success of the fitted map in reproducing the
bifurcations of the period-8 solutions, however, stops here. The period-8
solutions
do not period-double (for the model) to a period-16 and above, nor is there a
reverse cascade; they
persist until $R/R_c = 12.0136$, when they disappear
in a saddle-node bifurcation with
the saddle period-8s  (compared with the experimentally interpolated
value of $ R/R_c \approx 12.006$).
The ANN was somewhat --but not completely-- successful in ``learning"
these bifurcations:  it can reproduce
a part of the transitions required to get to the period-doubling,
but it does not ``go all the way" (the eigenvalues do not get to cross $-1$).
This shortcoming may be at least in part attributed to the fact that there is
a comparatively small amount of  experimental (training) data in this regime,
and that no transient data were available; nevertheless,
as we will discuss below,
there is a ``quick fix" for this problem.

\begin{figure}[p]
%\centerline{\psfig{file=./figures/figur13b.eps,height=5in}}
\vspace{5in}
\caption{Representative phase portraits predicted by the first ANN trained
(low $R/R_c$). The parameter values ($R/R_c$) of the model for the
portraits are as follows: A: 11.9430, B: 11:9513, C:11.9541, D: 11.9653,
E: 11.9848, F: 11.9876, G: 11.9898, H: 11.9374, I:11.9485, J:11.9530,
K: 11.9653, L: 11:9745, M: 11.9775, N: 11.9792, O: 11.9876 and P: 11.9898.}
\label{fig:lowportraits}
\end{figure}

This initial ANN is significantly more successful in capturing the low-R
bifurcation patterns (see Fig.~\ref{fig:lowportraits}):
a ``reverse" global bifurcation, involving again the
stable and unstable manifolds of the saddle-period 8 solutions gives rise to
two finite-amplitude invariant circles at $R/R_c \approx 12.0151$. After the
period 8 solutions have disappeared in a saddle-node bifurcation, closing the
period-8 isola, the only visible attractor left is these finite-amplitude
period 2 invariant circles. Of course, the ANN predicts both period-2 periodic
solutions: the saddle, and the source whose projection is located ``inside"
the two invariant circles \footnote{ Since the map
is 4-dimensional, the ``source" is really a {\it saddle}: it has two stable
eigenvalues in the full space, and two unstable ones. We call it a ``source"
because of the analogy with resonances and Hopf bifurcations in maps {\it of
the
plane}.}. This is clearly seen in Fig.~\ref{fig:lowportraits}G.
As the Rayleigh number is decreased, the finite-amplitude period-2 invariant
circles continue to exist (Fig.~\ref{fig:lowportraits}F and E) but the ANN
correctly predicts hysteresis between them and a large amplitude attractor
(Fig.~\ref{fig:lowportraits}P). The appearance of this attractor is associated
with a global bifurcation involving the saddle period-2 solutions.
This can be seen in Fig.~\ref{fig:lowportraits}P by the proximity
of the attractor and the saddle period-2 solutions.  Detailed calculations
show that in the hysteretic regime, one side of the (one-dimensional) unstable
manifold of the saddle period-2 solutions asymptotically approaches the
finite amplitude period-2 invariant cirles, while the other side asymptotically
approaches the large amplitude attractor. At the high-R
``boundary" of the hysteresis
a homoclinic interaction occurs between the unstable and stable manifolds of
the
period-2 solutions. The quotation marks indicate that this transition does
not occur at a single parameter value, but through the onset and end of a
homoclinic tangle. This tangle would appear ``heteroclinic"
for the second iterate
of the Poincar\'e map: the stable manifold of
each of the period-2 points interacts
with the unstable manifold of the ``other" period-2 point.

Returning to the ``upper" solution branch,
as the Rayleigh number is decreased through the hysteresis regime (towards
Fig.~\ref{fig:lowportraits}E), the finite amplitude period-2 invariant
circles approach the period-2 saddles, and finally disappear. Their destruction
involves another global bifurcation: another homoclinic interaction of the
stable and unstable manifolds of the saddle period-2 points. The difference
is that now the ``other" sides of the one-dimensional unstable manifolds are
involved, and the bifurcation would also appear as homoclinic for the
{\it second} iterate of the map. The two boundaries of hysteresis constitute,
therefore, two different global bifurcations involving the saddle period-2
invariant manifolds.

Phase portraits \ref{fig:lowportraits}E and \ref{fig:lowportraits}N also
illustrate the need for embedding in a space with more than two
dimensions: in the projection used, one of the two period-2 invariant
circles falls ``inside" and the other one ``outside" the coexisting
large amplitude attractor. This would not
be possible for a  {\it uniquely invertible}
map of the plane.

As the Rayleigh number is further decreased
(Fig.~\ref{fig:lowportraits}N through H)
the large amplitude attractor continues to exist; it deforms as the
parameter changes, and develops a number of frequency lockings, the most
prominent of which is the period-5 solution visible in
Fig.~\ref{fig:lowportraits}I
(compare also with the experimental data in Fig.~\ref{fig:ps2_sequence}).

The ANN is therefore capable of qualitatively
capturing all the phase portraits observed
experimentally. In addition, now that we can calculate the saddle-type
solutions
and study numerically their unstable manifolds, we have confirmed the nature
of the global bifurcations we proposed in the previous section
as marking the boundaries of the
hysteresis between the finite-amplitude period-2 invariant circles and the
large amplitude attractor.

\begin{figure}[p]
%\centerline{\psfig{file=./figures/ec1-2fig14.ps,height=5.62in,width=6in}}
\vspace{5.62in}
\caption{Bifurcation diagram computed using the first ANN trained.
Solid lines indicate stable periodic solutions,
broken lines indicate unstable solutions and filled circles indicate
period-2 small amplitude invariant circles (small) or large amplitude
invariant circles (large). Notice the hysteresis around $R/R_c \approx 11.99$
and the isolated solutions around $R/R_c \approx 11.95$.}
\label{fig:lowdiagram}
\end{figure}

\begin{figure}[p]
%\centerline{\psfig{file=./figures/figure15b.eps,height=6in,width=2in}}
\vspace{6in}
\caption{Representative phase portraits predicted by the
second ANN trained
(high $R/R_c$). The parameter values ($R/R_c$) of the model for the
portraits are as follows: A: 11:9942, B: 12:0015, C: 12.0135, D: 12.0140,
E: 12.0149, F: 12.0179, G: 12.0213, H: 12.0218, I: 12.0223 and J: 12.0433.}
\label{fig:highportraits}
\end{figure}

No further hysteresis was observed in this regime during the experimental runs,
for either ascending or descending Rayleigh numbers.
The last element of our postulated bifurcation diagram to be confirmed was
the saddle-node (saddle-source) bifurcation marking the end of the
$2/13$ resonance horn.
As can be seen, however, in Fig.~\ref{fig:lowportraits}D through A, the ANN
predicts a new region of hysteresis (actually an isolated region).
A saddle-stable node bifurcation is predicted at $R/R_c = 11.9386$,
giving rise to
a {\it stable} period-2 attractor (Fig.~\ref{fig:lowportraits}A). This
attractor coexists with the large amplitude one
(Fig.~\ref{fig:lowportraits}I, J)
and the basin boundary consists of the stable manifolds of the saddle type
period-2 solutions formed in the same saddle-node bifurcation (the ``resonant"
saddle period-2 as we have been referring to it).
As R is increased, a Hopf bifurcation occurs at $R/R_c = 11.9484$
giving rise to two small
amplitude period-2 invariant circles. These grow in amplitude, and are lost
in one more global bifurcation(s) involving the saddle period-2
stable and unstable
manifolds, leaving the large-amplitude attractor as the only stable solution.
The nature of this global bifurcation is again obvious from
Fig.~\ref{fig:lowportraits}C and D
from the relative location of the period-2 stable invariant circles and the
resonant period-2 saddles.
The low-R predictions of this ANN are summarized in the
computed bifurcation diagram
of Fig.~\ref{fig:lowdiagram}, which illustrates the
hysteretic regime observed experimentally (period-2 invariant circles with
large amplitude attractor), as well as the yet unconfirmed
lower-R multistability region predicted by the ANN.

\begin{figure}[p]
%\centerline{\psfig{file=./figures/ec1-32bifall.ps,height=5.62in,width=6in}}
\vspace{5.62in}
\caption{Bifurcation diagram computed using the second ANN trained.
Solid lines indicate stable periodic solutions,
broken lines indicate unstable solutions and filled circles indicate
period-2 small amplitude invariant circles.}
\label{fig:highdiagram}
\end{figure}

\begin{figure}[t]
%\centerline{\psfig{file=./figures/ec1-32bifinset.ps,height=3.8in,width=4in}}
\vspace{3.8in}
\caption{Blowup of the previous bifurcation diagram in
the neighborhood of the period-8 isola. Notice the pair
of period doublings bifurcations to period-16 solutions. A forward and
a reverse cascade of
period doublings (not shown in detail) occur around $R/R_c \approx 12.02$.}
\label{fig:blow16}
\end{figure}

Since this additional multistability region is isolated, it is quite natural
that it was not observed with the experimental procedure followed. The
ANN therefore predicts a bifurcation sequence equally plausible (and equally
compatible with the data) as the one we initially guessed and illustrated
schematically in Fig.~\ref{fig:schemebif1}. While it confirms a number of
our ``guesses" on the bifurcations underlying the transitions between different
phase portraits, it also poses a new experimental question: how could we
confirm or rule out the predicted low-R multistability region?
In experimental terms,
how could we systematically try to obtain data on this isolated solution
branch? This is precisely the type of feedback we expect from a
data post-processing procedure, and we will further discuss  below the
design of experiments motivated from this prediction.

We now return to the single inconsistency between this ANN and
the experimental data: while it essentially
matched every experimental observation, the network did not predict
the forward and reverse cascade of period-doublings of the
period-8 solutions to apparent chaos. As we discussed above, this
should be at least partly attributed to the comparatively small
number of data available in this regime, and the lack of transient
information. To rectify this, we gave more weight to the high-R
experimental data in the overall training set (to be exact, we
considered the data of panels Q, R, S and T in Fig.~\ref{fig:all1}
and panels J, K and L in Fig.~\ref{fig:all2} twice in the training
set). We then expect this ``second" ANN to more accurately capture
the dynamics at higher values of R; this is indeed the case, as
can be seen in both the computed phase portraits of
Fig.~\ref{fig:highportraits}
and the corresponding computed bifurcation diagram of
Fig.~\ref{fig:highdiagram}.

Consider the sequence of representative phase portraits predicted by
the second ANN (Fig.~\ref{fig:highportraits}A through J). At high values of
R (Fig.~\ref{fig:highportraits}J) we find the stable period-2 resonant
solution,
which --as in the case of the first ANN-- undergoes a Hopf bifurcation to
a small-amplitude period-2 invariant circle (Fig.~\ref{fig:highportraits}I).
The sequence of bifurcations starts here exactly as was discussed above
for the first ANN: a saddle-node period-8 bifurcation, giving a stable
and a saddle period-8 (Fig.~\ref{fig:highportraits}H)
(along with a small interval of hysteresis, too narrow
to be observed experimentally); then loss of the small amplitude
invariant circle in
a global bifurcation involving the stable and unstable manifolds of the saddle
period-8 solutions. But now the sequence changes, and the second ANN
does indeed reproduce the experimental bifurcations: a period-doubling of
the period-8 to a period-16 solution ($R/R_c = 12.0223$,
see Fig.~\ref{fig:highportraits}G)
and a subsequent cascade of period doublings resulting in an 8-horseshoe
segment
apparently chaotic attractor (Fig.~\ref{fig:highportraits}F). This is followed
by a reverse cascade of period doublings ending
up with a period 16 (Fig.~\ref{fig:highportraits}E)
and a period-8 (Fig.~\ref{fig:highportraits}D) exactly as the
experimental data show.
The period-8 isola closes in a saddle-node bifurcation
($R/R_c = 12.0085$) just after two
finite-amplitude ``invariant circles" are formed by a homoclinic
interaction of
the saddle period-8 solutions (jump to the
attractor in Fig.~\ref{fig:highportraits}C after
a very narrow hysteresis interval). These finite-amplitude
period-2 invariant circles
continue to deform smoothly as R decreases further, and eventually are lost
around $R/R_c = 11.9940$. As we discussed in detail above,
this is due to a global bifurcation
(a homoclinic interaction) involving the stable and unstable manifolds of the
resonant saddle period-2 solutions. This is again
indicated in Fig.~\ref{fig:highportraits}A
by the proximity of the projections of the two invariant circles with
the resonant period-2 saddles
predicted by this network.
Exactly as we discussed above, this bifurcation
causes a ``jump" to the coexisting
large amplitude attractor. The sequence is summarized in
the computed bifurcation
diagram of Fig.~\ref{fig:highdiagram}, which --along with its blowup
in Fig.~\ref{fig:blow16}-- should be
compared with the schematic diagram we postulated in Fig.~\ref{fig:schemebif1}.
The qualitative agreement was excellent for lower values of R  and partially
successful in the high-R regime with the first ANN; with the second ANN the
agreement now extends over the entire region of available experimental data.

\section{Discussion}

We found that the use of ANNs to study our experimental data from the
quasiperiodic regime of Rayleigh-B\'{e}nard convection constitutes a valuable
post-processing tool. The deterministic maps fitted to Poincar\'{e} sections
of the experimental time-series measurements were, as a rule,  accurate
in reproducing the short term system dynamics. They were also quite
successful in predicting the long-term attractors, and providing
the sequence of bifurcations leading from one to the other.

\begin{figure}[p]
%\centerline{\psfig{file=./figures/rbschem2-2.eps,width=6in,height=5.6in}}
\vspace{5.6in}
\caption{Schematic one parameter bifurcation diagram, showing
a summary of the ANN predictions.
Solid lines indicate stable periodic solutions,
broken lines indicate unstable solutions and filled circles indicate
period-2 small amplitude invariant circles (small) or large amplitude
invariant circles (large).}
\label{fig:netpreschem}
\end{figure}

The one-parameter bifurcation diagram found using the ANN (summarized
schematically in Fig.~\ref{fig:netpreschem}) is consistent
with the data; it guides --through confirmation or by providing a possible
alternative-- our interpretation of the phase portraits and our guesses of the
transitions between them as the Rayleigh number varies. It also guides
our experimental search of critical parameter values, by interpolating
or extrapolating from the available data.
One particularly interesting feature of the bifurcation diagram predicted
by the ANN is a multistability regime which was not observed experimentally.
This regime, however, {\it could not} be observed experimentally the way the
experimental procedure was performed, since the ``additional" attractors were
isolated. This example provides an indication of how the ANN can guide a
subsequent experimental search: if a two-parameter diagram is constructed
(by training the ANN using data from additional one-parameter cuts at
different Rayleigh numbers) we can draw a good guess of a path in two-parameter
space that could take us smoothly to the isolated attractors. This path
(as can be deduced from the two-parameter picture postulated below)
would involve sequences of experiments at higher Prandtl numbers, when
the isolated attractor branch ``connects" with the attractors we did
trace experimentally.

\begin{figure}[p]
%\centerline{\psfig{file=./figures/rbtwoparam3.eps,width=4in,height=3.75in}}
\vspace{5.75in}
\caption{Schematic two parameter bifurcation diagram, showing
the 2/13 resonance horn in the neighborhood of the experimental
data. The nature of the one-parameter bifurcation curves can be
inferred from the short description in Table 1.
The middle one-parameter cut corresponds to the one
actually predicted by the second ANN, while the upper is the one
we postulated upon inspection of the experimental phase portraits.
The lower one-parameter cut is the one identified by the first ANN
trained.}
\label{fig:bigbifschem}
\end{figure}

\begin{figure}[p]
%\centerline{\psfig{file=./figures/twoporschem.ps,width=5in,height=6in}}
\vspace{6in}
\caption{Schematic phase portraits representative of the
correspondingly numbered regions of the postulated two-parameter
bifurcation diagram (Fig.19).}
\label{fig:bigbifschem1}
\end{figure}

\begin{figure}[p]
%\centerline{\psfig{file=./figures/twoparschem2.ps,width=5in,height=5in}}
\vspace{5in}
\caption{Schematic phase portraits (a) and blowups (b) representative of the
regions numbered 7 and 11 in the postulated two-parameter
bifurcation diagram (Fig.19). Notice the hysteresis between invariant
circle period-2 solutions and period-8 fixed points, and the difference
in the relative arrangement of the coexisting attractors and the
period-8 saddle invariant manifolds.}
\label{fig:bigbifschem2}
\end{figure}

A significant stumbling block in this work was the determination of the
dimension of the embedding space. While all the phenomena predicted
by the network could in principle occur in maps of the plane, we could not
locate a simple two-dimensional Poincar\'{e} section of the data that
would give plausible phase portraits. The fact, for example, that one of
the two small period-2 invariant circles seems to fall ``inside" the
coexisting large amplitude invariant circle in Fig.~\ref{fig:lowportraits}P,
while the second period-2 invariant circle seems to fall ``outside",
is an indication that folds exist in this projection. We had to use
a higher number of delays in order to obtain such a picture with
a deterministic invertible map. In principle there is no problem with
a higher dimensional map (the embedding dimension based on Takens' version
of Whitney's theorem {\it should} be higher than two), but working in
a four-dimensional space prevents us from visualizing the stable manifolds
of the saddle resonant solutions, which now are three-dimensional
hypersurfaces in four-space. Had we been able to find a convenient projection
that would allow us to fit a map of the plane, it would have been possible
to numerically approximate the stable manifolds of the saddle-type solutions
too. The procedure used here should in principle be applicable to data and
attractors of higher dimensions; it is, however, important for the success of
the process that extensive data are available not only on the
attractor -- and,
for that matter, distributed over the entire attractor in phase space --,
but also on transients approaching it. In such cases the number of data
points used in training would grow significantly, and then the
massively parallel algorithm we used would be truly valuable. For the
case studied here (approximately 15000 data points) it is still possible
to train the ANN on a powerful workstation.

Another point worth making is the invertibility of the fitted map
(i.e. that every point should possess a unique preiterate, or, in
other words, trajectories on the attractor do not cross forward or
backward in time). This is a fundamental property of Poincar\'{e} maps
obtained from sets of ODEs, and hence it is a natural constraint on the
process. In principle, however, neural network maps (like polynomial or
rational polynomial maps) are not necessarily uniquely invertible; there
may exist regions of phase space for which several preimages of a given
phase point are possible. This should not be thought of as an insurmountable
problem: a little thought (and a study of the 1989 paper by E. Lorenz
\cite{lorenz}) will
show that every simple explicit integration scheme can be noninvertible!
In our case, the requirement is that the ANN
should remain uniquely invertible
{\it in the relevant regions of phase space}: those corresponding to
realistic initial conditions, and {\it especially} in regions containing
training data. The phase portraits predicted by the ANN will be consistent
with those predicted by deterministic
and uniquely invertible maps as long as initial conditions in the
``relevant" region have a single preimage
--or, if they have more than one, only one falls in the ``relevant"
region while all the additional spurious ones are far
away (in phase space) from the experimental data.
This invertibility problem can also be
avoided if we perform an identification of a continuous time system
(a set of ODEs) as opposed to a discrete map \cite{kevrekidisNBT}; but then
the geometric simplification of the Poincar\'e section is lost.

With some weighting of the data in one regime of initial discrepancy, we have
now (a) a complete qualitative -and reasonable
quantitative- consistency between
the experimental and the predicted data, (b) a confirmation of the local and
global bifurcations underlying the observed transitions
and hysteresis regimes, and
(c) a guide to further design of experiments.

Fig.~\ref{fig:bigbifschem} is a plausible schematic two-parameter bifurcation
diagram, including
what would correspond to both Rayleigh and Prandtl number changes. While in
this
paper we quantitatively studied data along a single one-parameter cut, detailed
two parameter experimental data as well as extensive studies of model systems
in
the quasiperiodic regime lead us to draw the interior of
the $2/13$ resonance horn
as shown. Table 1 summarizes the various bifurcations
in Fig.~\ref{fig:bigbifschem} in
a compact form, while Fig.~\ref{fig:bigbifschem1} and
Fig.~\ref{fig:bigbifschem2}
provides a quick schematic reference
of the qualitative phase portraits expected in each region.

The sides of the resonance horn are saddle-node
bifurcation curves of period-2 solutions.
A secondary Hopf bifurcation curve arises at a postulated
Takens-Bogdanov point on
the left hand side of the resonance horn. This
is a point at which two eigenvalues of the
Poincar\'e map are equal to 1, and one expects both a secondary Hopf
bifurcation
curve (a Hopf of the resonant period-2 solutions to period-2 invariant circles)
as well as a global bifurcation involving these
invariant circles {\it and} the saddle
resonant period-2 solutions, in the neighborhood of such a point.
A ``downward" dip of the secondary Hopf curve is consistent with experimental
data at lower values of 1/$\sigma$ (\cite{mainieriPRL1,eckePLA}).
A different homoclinic global bifurcation, associated
with the hysteresis between the
resonant period-2 and the large amplitude attractor, ``starts" lower down on
the
left-hand-side of the resonance horn. This type
of bifurcation is discussed in detail
in \cite{aronson,eckePLA}.
The secondary Hopf bifurcation has its own resonance horns, one of which
(the one giving rise to a period-8 solution) is shown schematically in
Fig.~\ref{fig:bigbifschem}.
This resonance horn --higher up in it-- contains the
successive nested period-doubling
curves, as well as --at its sides-- the narrow hysteresis
regions with the coexistence
of the period-8 solutions and the small amplitude period-2 invariant circles.

The  sequence of bifurcations observed in the experimental
one-parameter cut can be
succintly followed with the help of Table 1.
If a one-parameter cut is taken schematically just a
little higher in 1/$\sigma$,
we obtain the bifurcation diagram postulated initially in Section 4.
If, on the other hand, a one-parameter cut is taken schematically
just a little lower in 1/$\sigma$, we obtain the bifurcation diagram
predicted by the first ANN trained: the one where all transitions
except the period doubling cascades of the period-8 solution were captured.
It is well known that two parameters are necessary to map the
quasiperiodic regime
(at least close to a Hopf bifurcation).
Even though our one-parameter dependent ANN was qualitatively quite successful,
this discussion clearly indicates that
two-parameter data are required in order to construct  a complete picture of
the bifurcations of the system.
This is a direction we are currently pursuing.

\begin{table} [t]
\centerline{ \begin{tabular}{||l|l||} \hline \hline
\multicolumn{2}{|c|}{TABLE 1} \\ \hline
Transition Boundary & \multicolumn{1}{c||}{Description of transition}\\ \hline
1-2 & Saddle-node (sink) bifurcation of the period-2 solution \\
1-4 & Saddle-node (source) bifurcation of the period-2 solution\\
2-3 & Hopf bifurcation of the period-2 solution \\
2-5 & {\it Global bifurcation,} homoclinic tangle of the manifolds \\
 & of the saddle period-2; associated with the formation \\
 & (or destruction) of a stable large amplitude invariant circle \\
3-4 & {\it Global bifurcation,} homoclinic tangle of the manifolds \\
 & of the saddle period-2; associated with the formation  \\
 & (or destruction) of a stable period-2 invariant circle \\
3-6 & {\it Global bifurcation,} homoclinic tangle of the manifolds \\
 & of the saddle period-2; associated with the formation \\
 & (or destruction) of a stable large amplitude invariant circle \\
5-6 & Hopf bifurcation of the period-2 solution \\
5-12 & Hopf bifurcation of the period-2 solution \\
5-13 & Saddle-node bifurcation of the period-2 solution \\
6-7 & Saddle-node bifurcation of the period-8 solution \\
6-8 & Saddle-node bifurcation of the period-8 solution \\
7-8 &  {\it Global bifurcation,} homoclinic tangle of the manifolds \\
 & of the saddle period-8; associated with the formation \\
 & (or destruction) of a stable period-2 invariant circle \\
8-9 & Period doubling bifurcation of the period-8 solution \\
8-11 & {\it Global bifurcation,} homoclinic tangle of the manifolds \\
 & of the saddle period-8; associated with the formation \\
 & (or destruction) of a stable period-2 invariant circle \\
8-12 & Saddle-node bifurcation of the period-8 solution \\
9-10 ... & Subsequent period doubling bifurcations\\
11-12 & Saddle-node bifurcation of the period-8 solution \\ \hline \hline
\end{tabular} }
\caption {Transitions associated with the two-parameter bifurcation
diagram.}
\end{table}

\section{Acknowledgements}

	This work was partially supported by DARPA/ONR grant
N00014-91-J-1850 (IK, RR, AL, RF), the Exxon Education Foundation (IK,RR),
and by the U.S. Department of Energy,
Office of Basic Energy Science, Division of Materials Science (RE).
IK also gratefully acknowledges a David and Lucile Packard Foundation
Fellowship.
\newpage

\appendix

\section{Experimental portraits used during the training of the ANNs.}

Figs. \ref{fig:all1} and \ref{fig:all2} show all the experimental data
used during the training of the ANNs.

\begin{figure}[p]
%\centerline{\psfig{file=./figures/allup.ps,height=5.45in,width=6in}}
\vspace{5.45in}
\caption{Experimental Poincar\'e sections, used for training of the ANN;
the data in this figure were obtained starting at low $R/R_c$,
and increasing with constant step.
The parameter values ($R/R_c$) are as
follows A: 11.941, B: 11.946, C: 11.950, D: 11.954, E: 11.958,
F: 11.962, G: 11.966, H: 11.970, I: 11.975, J: 11.979, K: 11.983,
L: 11.987, M: 11.991, N: 11.995, O: 12.000, P: 12.004, Q: 12.008,
R: 12.012, S: 12.016, T: 12.020, U: 12.025, V: 12.029,
W: 12.030, X: 12.037 Y: 12.041, Z: 12.046 and A$_2$: 12.050.}
\label{fig:all1}
\end{figure}

\begin{figure}[p]
%\centerline{\psfig{file=./figures/alldown.ps,height=5.45in,width=6in}}
\vspace{5.45in}
\caption{Experimental Poincar\'e sections, used for training of the ANN;
the data in this figure were obtained starting at high $R/R_c$,
and decreasing with constant step.
The parameter values ($R/R_c$) are as
follows A: 12.053, B: 12.048, C: 12.044, D: 12.040, E: 12.036,
F: 12.032, G: 12.028, H: 12.023, I: 12.019, J: 12.015, K: 12.011,
L: 12.007, M: 12.003, N: 11.998, O: 11.994, P: 11.990, Q: 11.986,
R: 11.982, S: 11.977, T: 11.973, U: 11.969, V: 11.965,
W: 11.961, X: 11.957, Y: 11.952, Z: 11.948, A$_2$: 11.944 and
B$_2$: 11.940.}
\label{fig:all2}
\end{figure}

\end{document}